%% file: main.tex
\documentclass[sigconf]{acmart}

\AtBeginDocument{%
  }

\setcopyright{acmcopyright}
\copyrightyear{2018}
\acmYear{2018}
\acmDOI{XXXXXXX.XXXXXXX}

\acmConference[Conference acronym 'XX]{Make sure to enter the correct
  conference title from your rights confirmation emai}{June 03--05,
  2018}{Woodstock, NY}
\acmPrice{15.00}
\acmISBN{978-1-4503-XXXX-X/18/06}

\usepackage{subfigure}
\usepackage{makecell}

\usepackage{enumitem}
\setitemize{noitemsep,topsep=0pt,parsep=0pt,partopsep=0pt}

\usepackage{forloop}

\newsavebox\MyBreakChar%
\sbox\MyBreakChar{}
\newsavebox\MySpaceBreakChar%
\makeatletter%
\newcommand*{\BreakableChar}[1][\MyBreakChar]{%
  \leavevmode%
  \discretionary{\usebox#1}{}{}%
}%
\makeatother

\newcounter{index}%
\newcommand{\AddBreakableChars}[1]{%
  \StrLen{#1 }[\stringLength]%
  \forloop[1]{index}{1}{\value{index}<\stringLength}{%
    \StrChar{#1}{\value{index}}[\currentLetter]%
    \IfStrEqCase{\currentLetter}{%
        {*}{\currentLetter\BreakableChar[\MyBreakChar]}%
        {/}{\currentLetter\BreakableChar[\MyBreakChar]}%
        {+}{\currentLetter\BreakableChar[\MyBreakChar]}%
        {\&}{\currentLetter\BreakableChar[\MyBreakChar]}%
    }[\currentLetter]%
  }%
}%

\def\arrow{\includegraphics[height=0.65em]{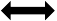}}
\def\navy{\includegraphics[height=1.0em]{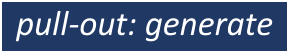}}
\def\yellow{\includegraphics[height=1.0em]{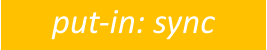}}

\def\vrT{\textsc{VR}}
\def\separatedT{\textsc{Separated}}
\def\unifiedT{\textsc{Unified}}

\def\vrB{\raisebox{-0.2\height}{\includegraphics[height=1em]{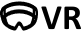}}}
\def\separatedB{\raisebox{-0.25\height}{\includegraphics[height=1em]{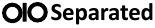}}}
\def\unifiedB{\raisebox{-0.2\height}{\includegraphics[height=1em]{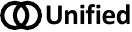}}}

\begin{document}

\title[]{Investigating Seamless Transitions Between Immersive Computational Notebooks and Embodied Data Interactions}

\author{Sungwon In}\orcid{0000-0002-5316-2922}
\affiliation{%
  \institution{Virginia Tech}
  \city{Blacksburg}
  \country{United States}}
\email{sungwoni@vt.edu}

\author{Eric Krokos}\orcid{0000-0003-1350-5297}
\affiliation{%
  \institution{Department of Defense}
  \country{United States}}

\author{Kirsten Whitley}\orcid{0000-0003-1356-326X}
\affiliation{%
  \institution{Department of Defense}
  \country{United States}}

\author{Chris North}\orcid{0000-0002-8786-7103}
\affiliation{%
  \institution{Virginia Tech}
  \city{Blacksburg}
  \country{United States}}
\email{north@cs.vt.edu}

\author{Yalong Yang}\orcid{0000-0001-9414-9911}
\affiliation{%
  \institution{Georgia Tech}
  \city{Atlanta}
  \country{United States}}
\email{yalong.yang@gatech.edu}


\renewcommand{\shortauthors}{Trovato et al.}


\input{body/00-Abstract}



\begin{teaserfigure}
  \centering
  \includegraphics[width=0.75\linewidth]{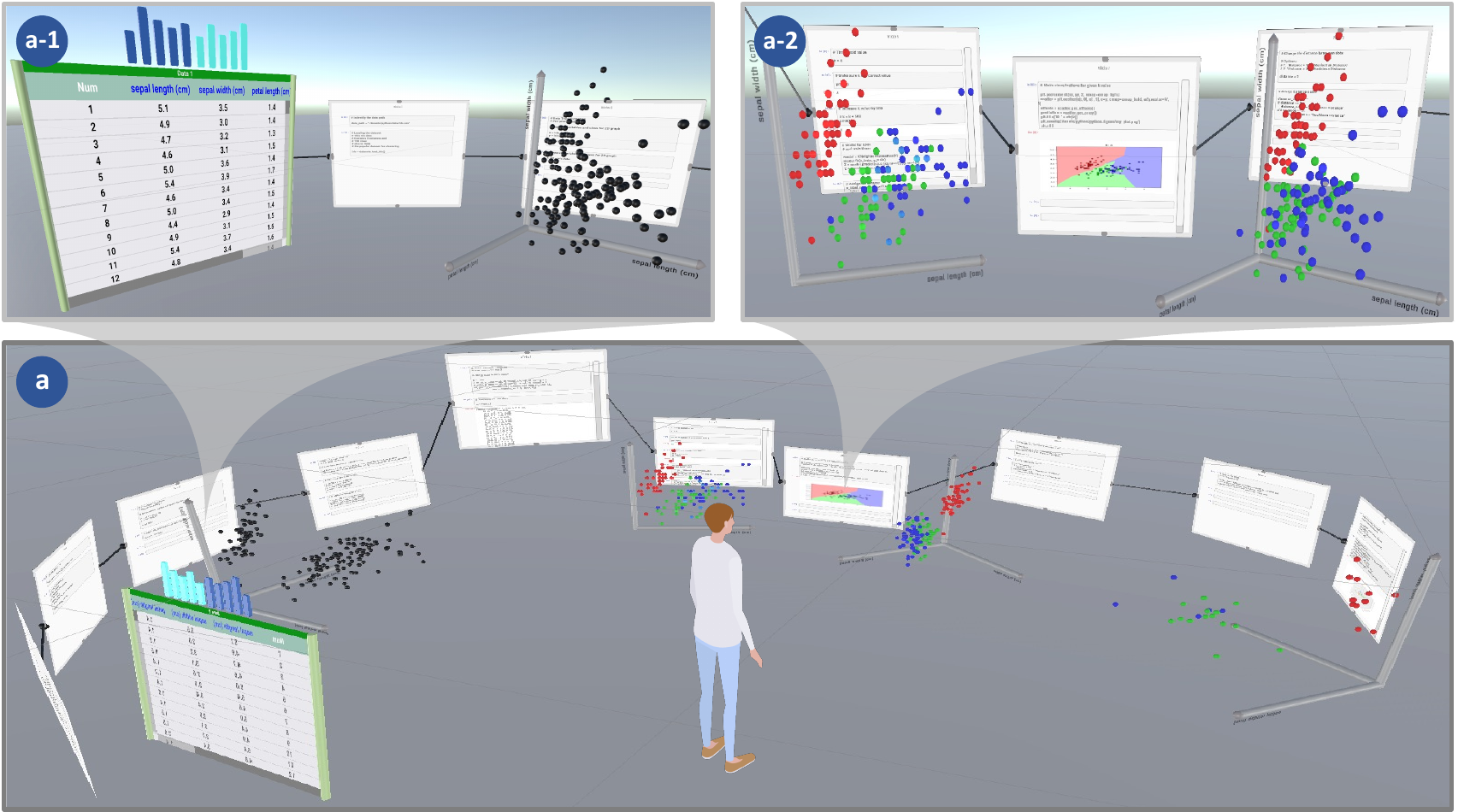}
  \caption{User performing data science tasks with ICoN. Within a unified environment, users can seamlessly transition between the notebook and embodied data interaction workspaces: (a-1) pulling out data tables and creating visualizations from the data table, (a-2) extracting and manipulating embedded visualizations. After exploring data tables and visualizations, the user can put them back into the computational notebook to update the code.}
  \vspace{1em}
  \label{fig:teaser}
\end{teaserfigure}

\maketitle

\input{body/01-Introduction-v2}
\input{body/02-Related_Work}

\input{body/03-Design_Goals-v2}

\input{body/04-Implementation}

\input{body/05-Evaluation_Study}

\input{body/06-Results}

\input{body/07-Discussion}

\input{body/09-Conclusion}


\begin{acks}
This research was supported by industry, government, and institute members of the NSF SHREC Center, which was founded in the IUCRC program of the National Science Foundation, and NSF grant IIS-2441310.
\end{acks}

\bibliographystyle{ACM-Reference-Format}
\bibliography{main}

\end{document}

%% file: body/00-Abstract.tex
\begin{abstract}
A growing interest in Immersive Analytics (IA) has led to the extension of computational notebooks (e.g., Jupyter Notebook) into an immersive environment to enhance analytical workflows.  
However, existing solutions rely on the WIMP (windows, icons, menus, pointer) metaphor, which remains impractical for complex data exploration.
Although embodied interaction offers a more intuitive alternative, immersive computational notebooks and embodied data exploration systems are implemented as standalone tools. 
This separation requires analysts to invest considerable effort to transition from one environment to an entirely different one during analytical workflows.
To address this, we introduce ICoN, a prototype that facilitates a seamless transition between computational notebooks and embodied data explorations within a unified, fully immersive environment.
Our findings reveal that unification improves transition efficiency and intuitiveness during analytical workflows, highlighting its potential for seamless data analysis.
\end{abstract}

%% file: body/01-Introduction-v2.tex
\section{Introduction}

Immersive Analytics (IA) is an emerging and promising field that aims to enhance data analysis processes through immersive technologies~\cite{ens2021grand,marriott2018immersive,zhao2022metaverse}.
Despite its potential, IA often focuses on individual components of data science and does not support the iterative nature of data science workflows well~\cite{wickham2016r}. 
Therefore, the current IA ecosystem often requires frequent switching between spatially disconnected environments (e.g., between separate tools).

One potential approach lies in computational notebooks, which have become a mainstream tool in data science due to their ability to integrate multiple tasks, such as data transformation, analysis, and visualization, into a single cohesive platform~\cite{chattopadhyay2020s}. 
Motivated by the success of computational notebooks, In et al.~\cite{in2024evaluating} tested a basic adaptation of computational notebooks in immersive environments.
Embodied interaction within immersive computational notebooks offers clear advantages for navigation, yet the WIMP (Windows, Icons, Menus, Pointer) paradigm is impractical when writing substantial amounts of code to perform data science tasks in immersive settings. 
In practice, analysts often rely on embodied approaches, but these introduce significant context-switching overhead, as users must alternate between immersive notebooks and external embodied data interaction tools, such as data transformation~\cite{in2023table} or visualization~\cite{cordeil2019iatk,zhu2024compositingvis}. 
In response to these limitations, we envision integrating embodied data interactions directly into immersive computational notebooks, thereby reducing context-switching costs and supporting a more seamless analytical workflow.

However, integrating embodied data interactions into immersive computational notebooks presents a key challenge, as it requires redefining how analysts manage multi-tasking workflows that involve both code execution and embodied data exploration in immersive environments. 
One common strategy of managing multi-tasking workflows in immersive analytics is to divide the available space into multiple sections, with each section dedicated to a specific task, to help maintain focus~\cite{in2025exploring}.
Inspired by this approach, we initially introduced two different workspaces, each placed in a separate environment: one for computational notebooks, where users execute the code, and another for embodied data interaction, where users manipulate data using embodied interactions.

Our next goal was to unify computational notebooks and embodied data interaction workspaces in a single environment, as transitioning within spatially separated environments introduces high context switching costs and disrupts the continuity of the analytical flow~\cite{yang2020embodied}.
However, maintaining seamless transitions in these unifications presents additional challenges, as computational notebooks and embodied data interaction workspaces differ in two fundamental ways: 
\textit{1. Dimensions of the artifact}: Notebooks primarily rely on 2D interfaces, while embodied data interaction workspaces involve 3D spatial artifacts; 
\textit{2. Interaction}: Notebooks operated with WIMP interactions, whereas embodied data interaction workspaces use gesture-based and physical movements.

To address this, we utilized embodied interactions to facilitate transitions between immersive computational notebooks and an embodied data interaction workspace (e.g., data transformation, visualizations).
We also extended embodied interactions to enable seamless transitions within the embodied data interaction workspace itself, such as between data transformation and visualization.
To this end, we propose a prototype system, ``ICoN'' (Immersive Computational Notebook), which utilizes embodied interaction throughout the entire data science pipeline within a single environment.

Unifying computational notebooks and embodied data interaction workspaces into a single environment may reduce context-switching and streamline analysis. 
However, such unification may also introduce visual clutter as the number of artifacts grows.
Conversely, a separated environment provides more space but increases switching effort. 
To systematically examine these trade-offs, we designed a user study to evaluate analysts' performance using two environment configurations, separated and unified.
Findings suggest that a unified environment facilitates intuitive transitions and improves workspace management in immersive environments.

In summary, our main contributions are:
\begin{itemize}[leftmargin=2.3em]    
    \item We designed and implemented ICoN, a research prototype system that enables seamless transitions between computational notebooks and embodied data interaction workspaces (e.g., data transformation and visualization), as well as within the embodied data interaction workspaces themselves.

    \item We conducted a controlled user study comparing a Unified (ICoN), where all workspaces exist in a single environment, and a Separated, where workspaces are distributed across spatially disconnected environments.
\end{itemize}

%% file: body/02-Related_Work.tex

\section{Related Work} 
\label{sec:relwork}

\textbf{Iterative Data Analysis Process.} 
In data science, analysts often transition between various stages within the iterative data analysis workflow~\cite{wickham2016r}, such as transitioning from data transformation to data visualization. 
These transitions are crucial for maintaining the continuity of the workflow.
However, analysts often rely on multiple separate tools and environments for different tasks, making transitions complex~\cite{yang2020embodied}.
While such transitions can be handled on desktops by switching between tasks through multiple tabs or windows, analysts in immersive analytics often need to close one application before opening another, a process that can easily break the sense of immersion~\cite{houben2017opportunities}.
Furthermore, there is a need to manage the compatibility of the data across different tools, which can easily introduce errors, inconsistencies, and inefficiencies~\cite{kery2018story, rule2018exploration}.
These challenges underscore the need for optimizing transitions to support an iterative data analysis workflow.

\textbf{Computational Notebooks.}
The concept of the computational notebook first emerged from Knuth's study of literate programming, aiming to create a computational narrative that seamlessly integrates visuals, text, and analytical insights within a single document~\cite{kluyver2016jupyter, peng2011reproducible, rule2018exploration}.
This approach has led to the widespread adoption of computational notebooks in data science as they enable a comprehensive analysis process, such as the leading application~\cite{JupyterSurvey}, Jupyter Notebook~\cite{JupyterSurvey, JupyterNotebook}.
Following this, various approaches have been explored, such as the non-linear representation in Observable~\cite{Observable} and Apache Zeppelin, focusing on managing big data~\cite{ApacheZepplin}.
However, despite various approaches, computational notebooks continue to present several challenges, such as high navigation costs and ineffective representations for comparative analysis~\cite{chattopadhyay2020s}.
In response to these limitations, In et al. introduced the concept of using computational notebooks in immersive spaces, where users can intuitively navigate through notebooks in a larger space~\cite{in2024evaluating}. 
They employed cells and outputs in multiple windows in a semi-circular layout, which are interconnected to form a complete notebook. 
While immersive computational notebooks significantly improve navigation performance, text input interaction using WIMP (Windows, Icons, Menus, Pointer) interfaces remains a challenge for immersive devices.
As a result, transforming and visualizing data within immersive computational notebooks continues to pose challenges for more complex and deeper explorations.

\textbf{Immersive Analytics.}
Immersive analytics leverages engaging and interactive technologies like VR and AR to significantly enhance data analysis by providing users with a more intuitive, embodied experience~\cite{ens2021grand, marriott2018immersive,zhao2022metaverse}. 
By enabling users to explore and interact with data in a physical space, immersive analytics enhances comprehension, encourages deeper insights, and facilitates a more intuitive understanding of complex datasets~\cite{cordeil2019iatk, cordeil2017imaxes,yang2025litforager,sicat2018dxr}. 
While tools like ImAxes offer a fully immersive visualization authoring experience, many existing solutions still require users to design and configure visualizations on a desktop before viewing them in an immersive setup.
Beyond visualizations, data transformation tools have been introduced for a broader audience by utilizing graphical user interfaces (GUI)~\cite{alteryx, tableauPrep, trifacta} or using gestures~\cite{jiang2013gesturequery, nandi2013gestural,nandi2013interactive}.
Notably, In et al. introduced immersive data transformation, enabling users to perform data cleaning and transformation within a fully embodied workspace, incorporating more visualizations and embodied gestures~\cite{in2023table}.
They found that embodied interaction, leveraging physical movements to engage with digital content in an embodied workspace, provides an intuitive and effective means of data exploration through direct manipulation~\cite{bach2017hologram, yang2020tilt,tong2025exploring,reiske2023multi}.

%% file: body/03-Design_Goals-v2.tex
\section{Design Rationale and Considerations}
\label{sec:design_goal}

Our goal is to integrate two distinct user interfaces and interaction paradigms in immersive environments to enhance data analysis. 
In immersive environments, spatially distributing computational notebook cells supports intuitive navigation, allowing analysts to move freely and quickly shift focus across various components~\cite{in2024evaluating}. 
However, pointing small targets and text input can disrupt user experience and flow~\cite{drucker_touchviz_2013}. 
\textit{Embodied interaction}, on the other hand, offers a promising alternative by enabling physical and direct manipulation of visual artifacts instead of traditional pointing and coding~\cite{in2023table,huang2023embodied}.
While embodied interactions have been successfully designed for individual data analysis tasks (e.g., data transformation and visualization), transitioning between and connecting immersive computational notebooks and embodied data interaction tools is challenging. 
Meanwhile, in desktop settings, file-based workflows connect multiple data analysis tools, but incorporating file management in immersive settings could introduce additional complexity and require precise interactions.

Given the strengths and limitations of both approaches, we propose a unified workspace to bridge computational notebooks and embodied data interaction.
However, simply merging them may introduce transition overhead that negates potential benefits. 
To guide our design, we draw from Csikszentmihalyi’s flow theory~\cite{czikszentmihalyi1990flow} to enhance fluidity in data analysis and Elmqvist et al.'s fluid interaction principles~\cite{elmqvist2011fluid} to ensure seamless transitions. 
Specifically, interactions should 1) promote flow, 2) support direct manipulation, and 3) minimize the gulf of action. 
Following these principles, we outline key design considerations to unify both systems to ensure a fluid and seamless transition experience.

\textbf{DC1: Provide low-effort transitions that promote flow.}
Broadly speaking, we envision two types of transitions in our explorations: \textit{navigational} and \textit{interactive}.
Navigational transition refers to changing the viewpoint, allowing analysts to access different information. 
In an immersive environment, natural walking or rotating the head is a standard and effective way to achieve navigational transitions~\cite{nilsson2018natural}. 
Interactive transition is a unique type of transition in our setting, involving the change of data artifact representations, for example, converting a static visualization image rendered in the computational notebook into an interactive visualization in the immersive space. 
Given the proven effectiveness of navigational transitions in immersive environments, we will focus on designing and building low-effort interactive transitions.

\begin{figure*}
    \centering
    \includegraphics[width=0.9\textwidth]{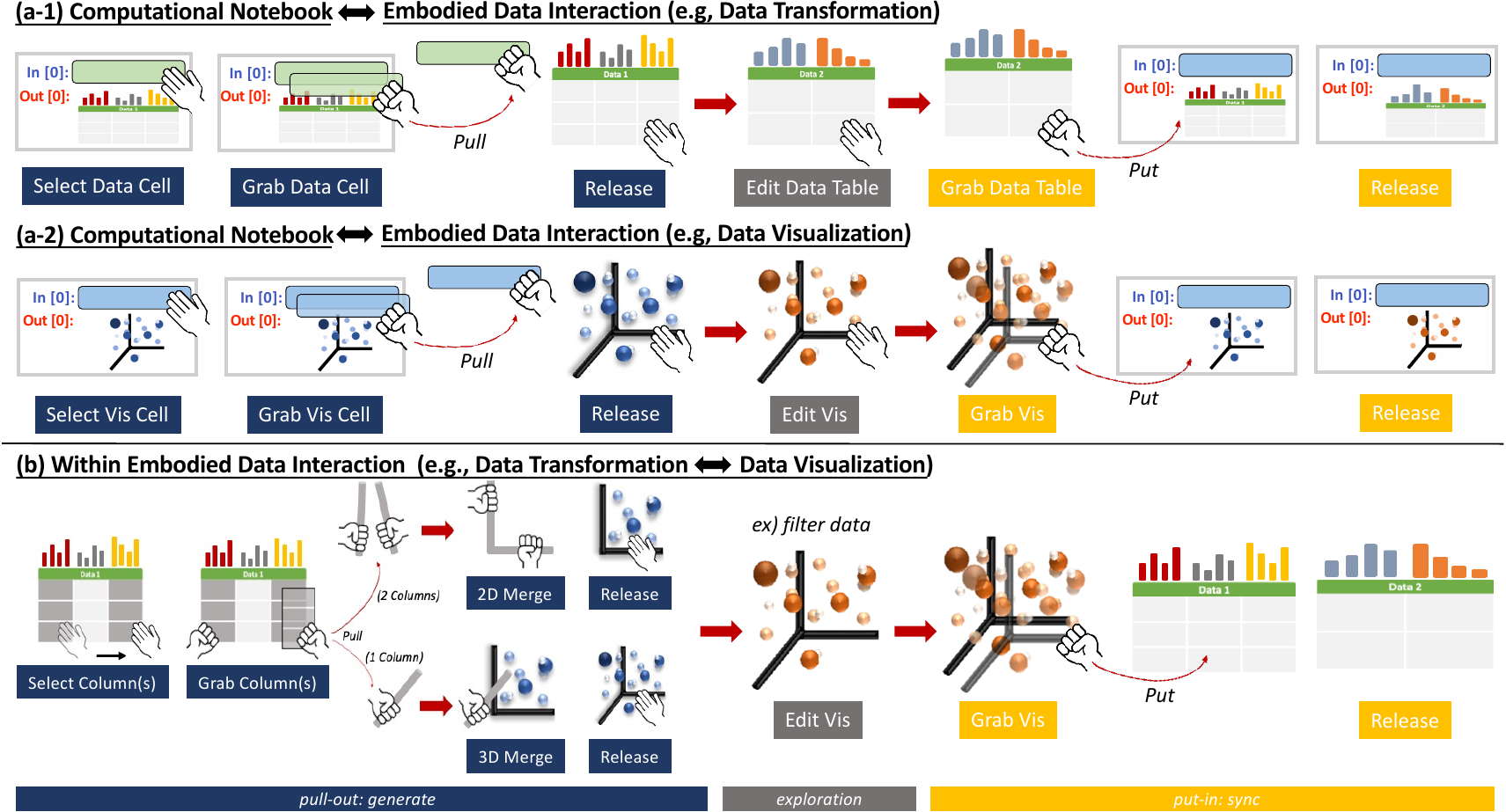}
    \caption{Gesture interactions to transit between computational notebooks and embodied data interaction (a-1 and a-2), and (b) within embodied data interaction.} 
    \label{fig:gesture}
\end{figure*}

\textbf{DC2: Enable embodied transitions that support direct manipulation.}
Direct manipulation refers to interacting with digital objects like interacting with physical objects~\cite{horak2020responsive, saket2017evaluating}. 
This approach allows the physical movement and the interacted object to exist in the same physical space, reducing spatial perception misalignment~\cite{yang2020tilt,huang2023embodied}. 
In immersive environments, one of the most common and well-recognized direct manipulations is grabbing and moving a digital object. 
We aim to leverage this intuitive interaction, particularly for interactive transitions where analysts frequently manipulate data rather than navigate.

\textbf{DC3: Provide intuitive and responsive embodied interactions to minimize the gulf of action.}
The gulf of action, comprising the gulf of execution and the gulf of evaluation, refers to the disconnect between user intentions and system responses~\cite{norman1986user}.
In immersive environments, this gap can be especially pronounced when transitioning between different states, tools, or modes.
To reduce the gulf of execution, we aim to design embodied interactions (e.g., gestures) that align closely with users' expectations for transitioning between different workspaces. 
For example, users might grab a chart and move it toward a computational notebook space to implicitly indicate an interactive transition.
To reduce the gulf of evaluation, the system should immediately reflect these transitions through responsive feedback, so users can easily perceive the new system state and confirm that their intention was recognized.

%% file: body/04-Implementation.tex
\section{ICoN}

Although computational notebook interfaces in immersive spaces are well defined~\cite{in2024evaluating}, they face a key challenge in supporting end-to-end data science workflows, as each component of immersive analytics~\cite{chattopadhyay2020s, in2023table, cordeil2019iatk} exists in an independent workspace. 
Specifically, while these workspaces allow analysts to focus more effectively on individual tasks, they also require the transition between spatially disconnected environments.
In response to these challenges, we developed ICoN, a prototype system that integrates embodied data interaction workspaces, specifically for data transformation and visualization, directly within immersive computational notebooks. 
Furthermore, ICoN supports seamless transitions between the computational notebook and the embodied data interaction workspaces, enabling a more fluid and unified analytic workflow.
The following discussion details our main design and implementations.

\label{sec:icon_design}

\subsection{Seamless transition}
\label{subsec: transitioning}
We prioritized designing interactions for interactive transitions, as slight shifts, rotations of the head, or eyes can accomplish navigational transitions.
We initially considered using teleportation to trigger interactive transitions, such that users could swipe to teleport to desired workspaces.
However, prior research shows that teleportation adds context-switching costs as users must adjust to dramatically changed environments~\cite{yang2020embodied}.
Through design, we found that pull-and-put interactions, facilitated by natural grabbing and walking, enable users to maintain focus while transitioning between workspaces.
We believe this approach fulfills our design consideration of providing low-effort transitions (DC1) and enabling embodied transitions (DC2).
Consequently, we employed pull-and-put interactions for interactive transitioning between computational notebooks and embodied data interaction workspaces. 

In addition to facilitating interactive transitions between the computational notebook and the embodied data interaction workspace, transitions also frequently occur within the embodied data interaction workspace~\cite{wickham2016r}, such as from data transformation to visualization tasks. 
To further support seamless interactive transitions, we extended the same pull-and-put interaction modalities to enable smooth transitions.
In the following sections, we detail the transition mechanism, particularly focusing on interactive transitions.

\textbf{Computational Notebook \arrow{} Embodied Data Interaction.}
We provided two types of interactive transitions between the computational notebook and embodied data interaction spaces, specifically focusing on data transformation and visualization.

Transition from \textit{computational notebook} to \textit{data transformation:} the user first needs to select a data cell (marked in green). 
The system identifies a data cell when it contains code for importing or assigning datasets.
Following this, the user grabs a selected cell and pulls it out.
Upon releasing it in the desired position, a data table will be generated, as illustrated in \autoref{fig:gesture}-(a-1)-\raisebox{-0.20\height}\navy{}.
As an indicator, the line will be generated between the data cell and the generated data table (see~\autoref{fig:vr_condition}-(b)).

Transition from \textit{data transformation} to \textit{computational notebook:} the user needs to grab the explored data table and put it back into the computational notebook, as illustrated in \autoref{fig:gesture}-(a-1)-\raisebox{-0.20\height}\yellow{}.
The explored data table will be recorded in the computational notebook, and the line indicator will be removed.
The details of how the code is recorded and executed are discussed in a later section.

Transition from \textit{computational notebook} to \textit{data visualization:} user first selects a visualization cell (marked in blue). 
Upon selection, users grab, pull out, and release the selected cell in the desired position.
The existing 2D or 3D static visualization image in the computational notebook will be generated in an interactive and three-dimensional manner, as illustrated in \autoref{fig:gesture}-(a-2)-\raisebox{-0.20\height}\navy{}.
As an indicator, the line will be generated between the visualization cell and the generated data visualization (see~\autoref{fig:vr_condition}-(b)).

Transition from \textit{data visualization} to \textit{computational notebook:} user grabs the explored visualization and puts it back into the computational notebook.
The line indicator will be removed, and the explored visualizations will be recorded in the computational notebook, as illustrated in \autoref{fig:gesture}-(a-2)-\raisebox{-0.20\height}\yellow{}.
We specify the code recording and executing process in a later section.

\textbf{Within Embodied Data Interaction.}
Transition from \textit{data transformation} to \textit{data visualization:} users select two columns, grab one with each hand, and merge them. 
The system then generates a 2D interactive visualization using the chosen columns, as illustrated in \autoref{fig:gesture}-(b)-\raisebox{-0.20\height}\navy{}.
As an indicator, the line will be generated between the data table and the visualization.
The selected columns will be marked in blue to indicate they have been used for visualization.
In addition, users can convert a 2D visualization into a 3D visualization by grabbing an additional column from the data table and adding it to an existing 2D visualization.
Users can also convert back to a 2D by discarding the desired axis.

Transition from \textit{data visualization} to \textit{data transformation:}  users grab the visualization and put it back into the data table, as illustrated in \autoref{fig:gesture}-(b)-\raisebox{-0.20\height}\yellow{}.
The data table will be updated with the explored visualizations, and the line indicator will be removed.

\subsection{Embodied Low-code programming}
The low-code programming approach~\cite{chattopadhyay2020s, in2024evaluating} is known for its rapid application development with minimal manual coding~\cite{drosos2020wrex, kery2020mage}.
We envision this approach further reduces the effort required for interactive transitions (DC1) and minimizes the gulf of actions (DC3).
Therefore, we enable the synchronization of data tables and visualizations into the computational notebook without writing extensive code.
While we opted for a low-code approach to further facilitate the interactive transition, we recognized that implementing a fully functional low-code system would be premature without an initial investigation of our transition methodologies.
Therefore, we focused on providing a low-code mechanism narrowed to specific tasks.
Below, we detail the adapted low-code mechanism in ICoN.

\textbf{Data artifacts generation mechanism in \raisebox{-0.20\height}\navy{}.} 
To generate data tables, we focused on using Python Lists, Arrays~\cite{PythonPackage}, and Pandas DataFrame~\cite{Pandas}. 
We first collect and save the required data (e.g., rows and columns) to the local machine (Unity) and then render data tables based on the stored data.
To generate data visualizations, we focused on the Matplotlib~\cite{MatplotLib} and Seaborn~\cite{Seaborn} packages. 
These packages allowed us to access the data needed for visualizations within the embodied data interaction workspace, such as the coordinates of data points, axis names, and clustering colors. 
We first ran the appropriate functions in Matplotlib or Seaborn (Python) to save the required data (e.g., array or list) into the local machine (Unity, C\#).
Then, we rendered visualizations by displaying data points with their clustered colors.

\begin{figure}
     \centering
    \includegraphics[width=1\columnwidth]{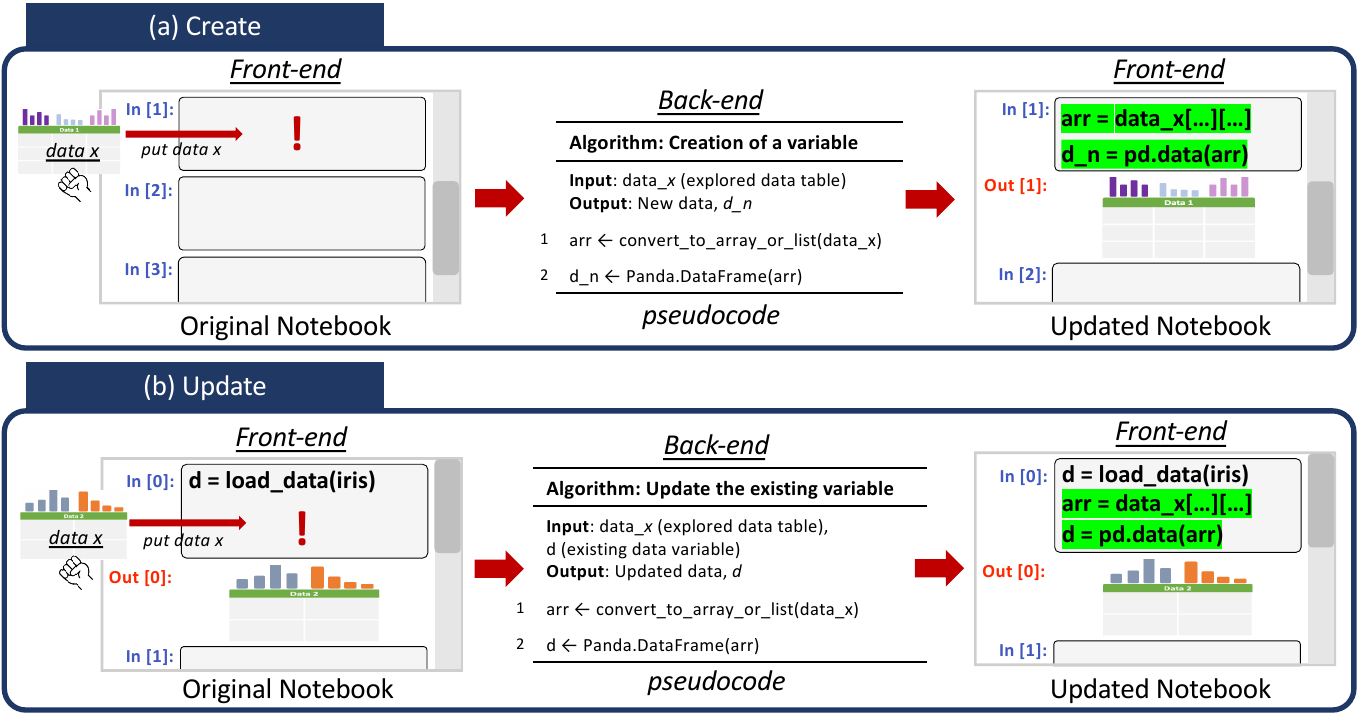}
    \caption{Illustrations of code generation mechanism with examples for creating and updating the variables. Users can either (a) create a new variable by putting their data into an empty cell, or (b) update an existing variable by putting their data back into a cell that was previously pulled.} 
    \label{fig:code_sync}
\end{figure}

\textbf{Code generation mechanism in \raisebox{-0.20\height}\yellow{}.} 
Users can update their explored data artifacts in the computational notebook by saving them as executable code.
The generated code is compatible with existing code cells and follows to two fundamental principles:
\textit{Create:} User can ``put-in'' their explorations into an empty cell.
The back-end system then converts the explored data into a compatible format, such as arrays or lists.
Based on the data type, the system generates a new variable, using a Pandas DataFrame for data tables or Matplotlib/Seaborn for data visualizations. 
The corresponding code is then displayed on the front-end, allowing users to easily follow and reflect on their actions, as shown in \autoref{fig:code_sync}-(a).
\textit{Update:} User can ``put-in'' their explorations into a cell previously pulled out to generate the data artifacts.
Similar to the logic used for creating variables, the back-end system converts the explored data into a compatible format, such as arrays or lists. 
However, instead of creating a new variable, it updates the existing variable based on the user's exploration. 
The updated code is then displayed on the front-end, enabling users to easily track and reflect on their actions, as shown in \autoref{fig:code_sync}-(b).

%% file: body/05-Evaluation_Study.tex
\section{User Study}
\label{sec:user_study}

While ICoN is designed to support seamless transitions, a unified environment can become cluttered as users interact with an increasing number of computational notebooks, data tables, and visualizations. 
In contrast, a separated environment reduces visual clutter but may increase the cognitive effort required to switch between spaces.
To investigate these trade-offs, we focus on the spatial placement of multiple workspaces as the primary factor.
Specifically, we compare a Unified environment, where the computational notebook and embodied data interaction workspace exist in the same environment, with a Separated environment, where participants can access only one workspace at a time.

In summary, we aim to answer two key research questions:
\begin{itemize}
    \item What are the trade-offs of transitions in an ICoN (unified) compared to separate workspaces to perform data science tasks within an immersive environment?
    
    \item What are the underlying factors of the trade-offs?
\end{itemize}

\subsection{Study Conditions}
\label{sec:user_study_conditions}
To support intuitive interaction, we used bare-hand input instead of physical devices. 
However, due to challenges with WIMP-based virtual keyboards in VR~\cite{in2024evaluating}, we provided a physical keyboard on a movable, height-adjustable desk. 
This allowed participants to navigate freely while accessing the keyboard as needed, minimizing impact on spatial exploration~\cite{davidson2022exploring}. 
The following describes our implementations for each experimental factor.

\separatedB\textbf{.}
We considered several organizational strategies, such as clustering and filtering. 
Clustering, which groups related items based on proximity or other attributes~\cite{fua1999hierarchical}, is a common technique for reducing clutter.
However, clustering risks disrupting the logical execution order by repositioning cells, which is an essential property of computational notebooks. 
Filtering, in contrast, allows users to temporarily hide artifacts while preserving their spatial positions~\cite{yang2003interactive}. 
This design also reflects the current ``one-world-at-a-time'' paradigm of XR knowledge work, where switching applications often requires fully exiting one context before entering another, leading to high context-switching costs~\cite{yang2020embodied}.

\begin{figure}
    \centering
    \includegraphics[width=0.83\columnwidth]{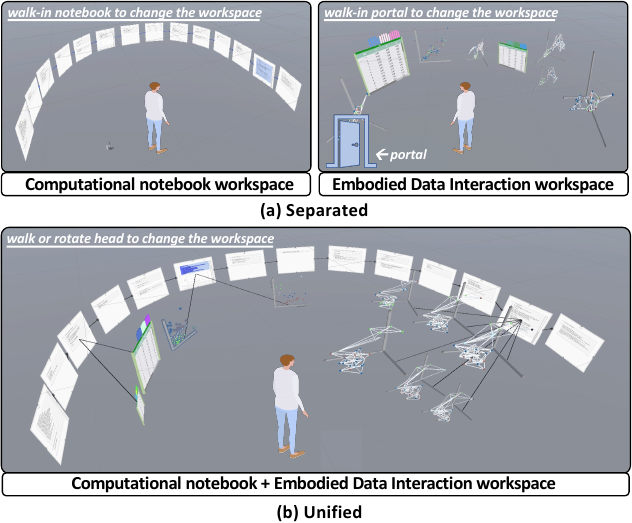}
    \caption{Illustration of tested conditions. 
    In  \separatedT, the notebook and embodied data interaction workspaces were spatially separated, requiring a walk-in portal to transition between them. In  \unifiedT, the workspaces were merged, enabling seamless transitions by walking or head rotation.}
    \label{fig:vr_condition}
\end{figure}

Based on this approach, in the \separatedT, our primary goal was to simulate current practices in immersive analytics, where only one computational notebook or data artifact is visible at a time. 
To achieve this, we implemented two spatially disconnected workspaces, where users could filter out or transition from one workspace to the other as needed.
Furthermore, we employed physical walking interactions to support these transitions, with participants transitioning between workspaces by walking into or out of workspaces.
Specifically, data artifacts were generated in front of participants when they entered a data or visualization cell, and they could return to the notebook space by walking through a portal positioned behind them (see \autoref{fig:vr_condition}-(a)). 
Interactive transitions required holding data artifacts, while navigational transitions could be performed without holding them.

\raisebox{-0.04\height}\unifiedB\textbf{.}
In the \unifiedT, the transition focus is on the user.
The computational notebook and data artifacts are visible at a time and spatially connected within a single environment (see \autoref{fig:vr_condition}-(b)). 
In contrast to \separatedT, physical arm interactions (put-in/pull-out) were utilized for interactive transitions in \unifiedT. 
Navigational transitions can be made by changing the viewpoint through head rotation and eye movement or by walking.

\begin{figure*}
    \centering
    \includegraphics[width=1\textwidth]{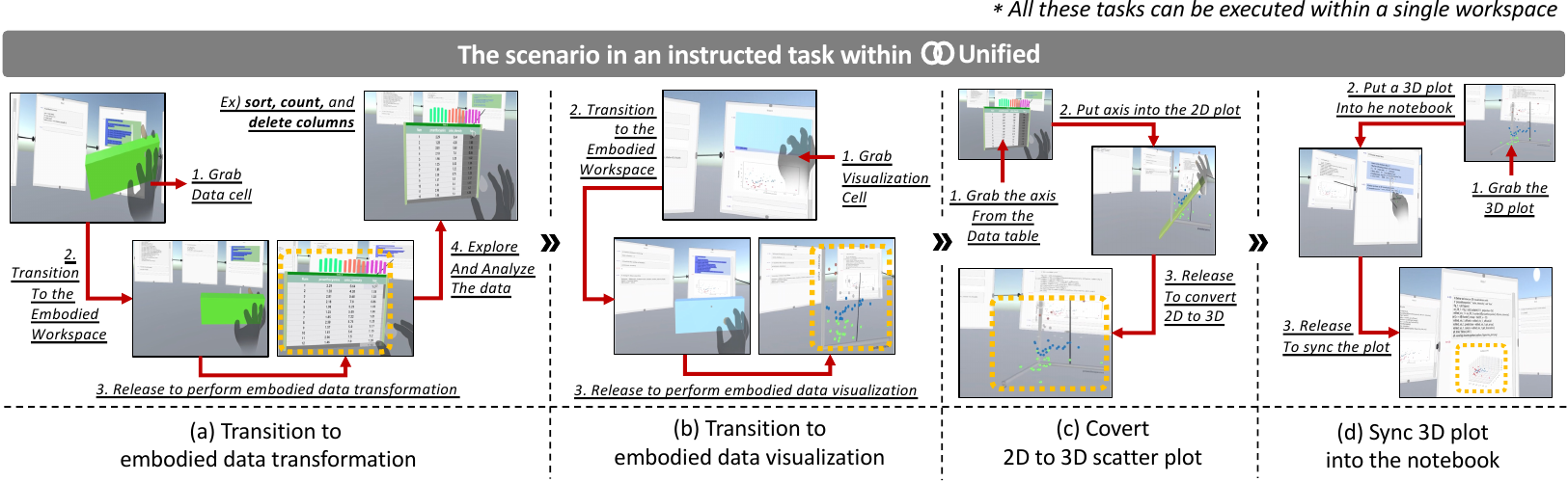}
    \caption{A scenario in an instructed task within \unifiedT. (a) The participant transitions to the embodied data transformation by generating a data table and exploring the data. (b) Transition to the embodied data visualization by generating a 2D scatter plot. (c) They grab a column from the data table and release it into the 2D scatter plot to convert it into a 3D scatter plot. (d) Finally, the participant synchronizes the 3D scatter plot with the computational notebook without writing code.}
    \label{fig:instruction}
\end{figure*}

\subsection{Data and Task} 
\label{sec:user_study_task}
Building on the initial evaluation of immersive computational notebooks~\cite{in2024evaluating}, with pre-existing code (14 windows, 30 cells) using two datasets from scikit-learn~\cite{scikit-learn}: wine and iris. 
A 2D scatter plot was provided for a brief overview of the data, while a 3D node-link diagram was utilized to visualize information from neighboring nodes to capture complex relationships.
Furthermore, we applied K-Means for clustering and K-Nearest Neighbors (KNN) for visualizing relationships in node-link diagrams, often referred to as KNN graphs that play a significant role in web-based search~\cite{bertier2010gossple} or recommendations~\cite{levandoski2012lars}. 
Based on this, we designed two tasks.

\textbf{1. Instructed task.}
In this task,  we provided step-by-step instructions to assess the interaction efficiency and ensure that the results were not influenced by other factors (e.g., the sense-making process).
The task began with participants understanding the dataset.
First, they generated a data table and reported the number of rows and columns. 
Next, they sorted to find a specific column's lowest and highest values. 
Participants were then required to document their analysis in the computational notebook.
Subsequently, they converted a 2D scatter plot to a 3D scatter plot by generating the 2D scatter plot from the computational notebook and adding a column from the data table to the existing plot.
The final step involved syncing the 3D scatter plot into the computational notebook.
The visual description of these instructions is illustrated in Fig. 6.

\textbf{2. Exploratory task.}
Following the instructed task, we introduced an exploratory task without specific instructions.
They were informed that the goal was to remove outliers and fine-tune the K-Means and KNN to perform simple data analysis with a 3D node-link diagram. 
While participants had the option to remove outliers manually by grabbing individual data points from the visualizations and discarding them, we instructed them to delete outliers by filtering specific columns in the data table. 
This ensured that outliers could be removed at once, reducing unnecessary effort in editing the data.
In addition, we informed participants that the column names and testing parameter value ranges were also available in the computational notebook.
Also, analyzing results from fine-tuning K-Means and KNN required the embodied data interaction workspace, as the computational notebook could not display all data points in the 3D node-link diagram.
Therefore, participants needed to iteratively transition between the computational notebook and the embodied data interaction workspace to remove outliers and analyze the results from different parameter values.
The task was completed when participants successfully deleted the outliers and identified the optimal parameter pairs.

\subsection{Participants and Apparatus}
We recruited 20 participants (16 male, 4 female, ages 18 to 35) from a university mailing list, selecting individuals based on their familiarity with data science, machine learning algorithms, and computational notebooks (e.g., Jupyter Notebook). 
Regarding VR experience, 14 participants had used VR, while the rest had no prior experience. 
For VR, we used the Meta Quest Pro headset with a resolution of $1800\times1920$. 
It was linked to a PC with an Intel i7-11800H processor and an NVIDIA GeForce RTX 3070 graphics card. 
Meta Air Link provided a cable-free experience, allowing participants to navigate a $25 m^2$ space freely. 
A movable desk with a keyboard was initially placed in front of them.

\subsection{Procedure}
Our user study followed a full-factorial within-subjects design, balanced using a Latin square.
After reviewing a consent form, participants adjusted their chairs and headsets for clear text visibility. 
Training sessions introduced computational notebook terminology and study conditions, with extensions to mitigate discomfort and learning curves. 
Training ended once participants demonstrated task proficiency (15–20 minutes). 
They then completed the study task, followed by a Likert-scale survey (System Usability Scale and NASA Task Load Index) and qualitative feedback on each condition. 
The study lasted an average of 90 minutes and concluded after completing all questionnaires.
No participants withdrew from the study or experienced cybersickness during the tasks.

\subsection{Hypothesis}
We formulated hypotheses based on prior explorations, and our analysis of the study conditions outlined in~\autoref{sec:user_study_conditions}.

\textbf{Error score.} 
Although participants could follow different paths to complete the tasks, the overall goals were clearly defined, and all participants successfully reached the correct solution without errors in the pilot study. Therefore, we did not expect differences in error rates across conditions.

\textbf{Transition in Unified and Separated.}
We anticipated that the \unifiedT~would facilitate more transitions compared to the \separatedT~($H_{space-trans}$) for both interactive and navigational transitions.
In the \unifiedT~, the computational notebook and data artifacts are visible simultaneously, enabling effortless transitions by simply moving the eyes or the head.
In contrast, the \separatedT~requires extra steps to change workspaces by walking in/out of the space.
We also anticipated that these additional steps would result in longer completion times ($H_{space-time}$).

\textbf{Workspace managements.}
We noted that interactive transitions naturally involve the creation of data artifacts.
As the analysis deepens and transitions become more frequent,
managing the growing number of artifacts within the given workspace becomes essential.
However, when analysts perceived the given space as unmanageable with the increasing number of artifacts, they tended to delete lower-priority items to avoid a messy environment~\cite{in2023table}.
This reduction in process history, often referred to as provenance~\cite{wang2015big}, would make it more difficult for analysts to track their prior processes.
As a result, we expect deleted more data artifacts and fewer data artifacts to be left in \unifiedT~compared to \separatedT~($H_{space-del\_left}$).
However, we expected that \unifiedT~would reduce mental demand ($H_{space-mental}$) by allowing artifacts to be displayed in proximity and spatially connected, which aids users in remembering the locations of previously visited cells/artifacts.

\begin{figure*}
    \centering
    \includegraphics[width=0.87\textwidth]{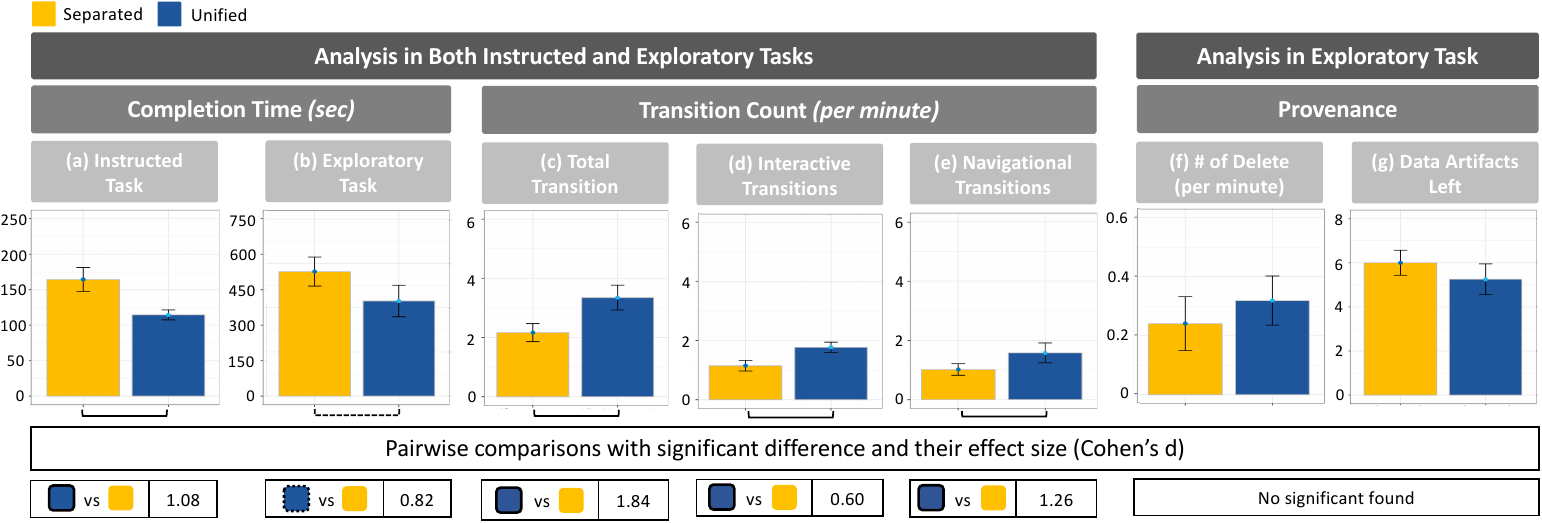}
    \caption{Collected quantitative results. (a) and (b) indicate the time taken to complete each task. (c) - (e) shows the number of transitions. (f) shows the number of deletions performed, and (g) the number of artifacts left. Solid lines indicate statistical significance with $p < 0.05$, and dashed lines indicate $p < 0.1$. Tables show effect sizes for pairwise comparisons, with black-bordered squares marking the better-performing conditions.} 
    \label{fig:performance}
\end{figure*}

\subsection{Measures}
We collected quantitative data and interaction records to evaluate our hypothesis. 
Specifically, we used the following measures;

\textbf{Error score:} an error was counted whenever a participant reported an incorrect answer.

\textbf{Time:} we measured the time from the initial exploration of the system to task completion. This directly relates to $H_{space-time}$, where time reflects overall efficiency and captures whether one condition allowed participants to complete organizational tasks more quickly than the other.

\textbf{Number of transitions:} we recorded the total number of transitions per minute to task completion. For navigational transitions, we counted whenever a participant changed the region they were focusing on. For interactive transitions, we counted whenever they attempted to manipulate code or data. For instance, when analysts switch their workspace to generate data from code, update code from data, or manipulate code/data immediately after changing their workspace.
This measure corresponds to $H_{space-trans}$, as it indicates how often participants switch their workspaces.

\textbf{Number of performed delete operations and data artifacts left:} we documented the number of data artifacts and the number of performed delete operations when the participant completed each study trial. This connects to $H_{space-del\_left}$, as it captures how participants managed workspace complexity, including whether they reduced clutter by discarding artifacts.

We also collected subjective ratings on a seven-point Likert scale for \textbf{mental demand}, \textbf{physical demand}, and \textbf{engagement}. 
Participants also \textbf{ranked} their overall experience.
\textbf{Qualitative feedback} about each condition's pros and cons was also collected.

%% file: body/06-Results.tex
\section{Results}
\label{sec:results}
This section reports the quantitative statistical analyses and summarizes the qualitative feedback for each condition.
We identified statistical significance at the following thresholds: $p < 0.001 (***)$, $p < 0.01 (**) $, $p < 0.05 (*)$, and $p < 0.1 (\cdot)$. 
In addition, we report the mean values with a 95\% confidence interval (CI) and use Cohen's d to measure the effect sizes of significant differences. 
Detailed statistical analysis results are provided in the supplementary materials.

\subsection{Quantitative Results}
The following sections present performance differences across the conditions.
Results are illustrated in ~\autoref{fig:performance} and~\autoref{fig:ratings}.

\textbf{Error score.} 
We noted that participants completed the study without significant trials, leading to no variation.

\textbf{Time.}
We measured the task's completion time to assess interaction efficiency across conditions. 
Our analysis showed that time was significantly affected by workspace configuration ($***$) for both computing environments. 
On average, within the instructed task, \unifiedT~was 49.9 sec (43.6\%) faster.
In addition, the \unifiedT~was consistently faster than the \separatedT~within the exploratory task, 133.1 sec (23.6\%) faster.
Therefore, we accept $H_{space-time}$.

\textbf{Number of transitions.}
Our findings indicate that \unifiedT~significantly impacted the both for navigational and interactive transitions ($***$).
We observed that for navigational transition, participants made 1.0 in \separatedT~and 1.6 in \unifiedT.
Furthermore, for interactive transition, participants made 1.5 in \separatedT~and 1.9 in \unifiedT.
Therefore, we accept $H_{space-trans}$ both for navigational and interactive transitions.

\textbf{Number of performed delete operations and data artifacts left.}
We observed no statistically significant difference in our tested task.
While we reject $H_{space-del\_left}$, our findings indicate that participants always deleted slightly more and kept less in \unifiedT~(0.31 and 5.25) compared to \separatedT~(0.24 and 6.00).

\textbf{Ratings and ranking.}
Significant effects were observed across mental demand ($***$), physical demand ($***$), and engagement ($***$).
\unifiedT~outperformed in terms of mental demand, physical demand (particularly in \vrT), and engagement.
Therefore, we accept $H_{space-mental}$.
In terms of engagement, \unifiedT~was the most engaging condition, with an average rating of 6.25 out of 7 and a confidence interval of 0.45.
In addition to subjective ratings, the \unifiedT~was the most preferred for performing data science tasks, with 15 participants ranking it first ($***$).

\subsection{Qualitative Feedback.}
We performed a qualitative analysis to identify recurring themes in user feedback across each condition. 
Two authors developed a coding scheme based on feedback from the first five participants, which was then consistently applied to all subsequent data. 
Finally, we synthesized the key insights across all conditions. Detailed coding results are provided in the supplemental materials.

\separatedB\textbf{.}
Participants favored \textit{flexible workspace management} (14), attributed to the \textit{large display space} (11). 
They reported that \textit{easy to analyze} (10) due to \textit{intuitive} (7) physical interactions and \textit{easy navigation} (6), which \textit{reduced mental load} (6). 
However, many participants expressed concerns about \textit{ineffective transitions} (10) caused by \textit{physical fatigue} (9). 

\unifiedB\textbf{.}
In contrast to \separatedT, participants reported \textit{effective transitions} (12) and \textit{preferred the unified workspace}~(8). 
They also reported that the \textit{large display space} (11), \textit{easy to analyze} (6), \textit{reduced mental load} (6), \textit{intuitive} (5) physical interaction, and \textit{flexible workspace management} (5). 
However, some participants expressed concerns about \textit{difficulties in managing the workspace} (4), noting that the \textit{limited workspace} (3) led to \textit{cluttered workspace} (3). 

%% file: body/07-Discussion.tex
\section{Key Findings, Observations, and Discussions}

\begin{figure}
    \centering
    \includegraphics[width=0.85\columnwidth]{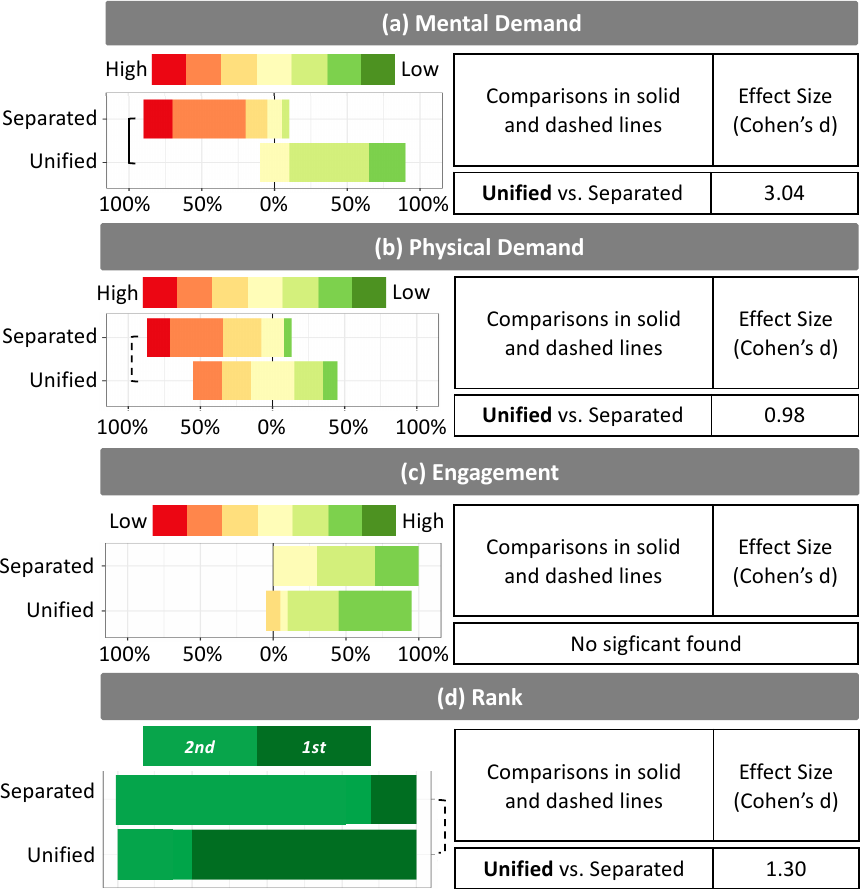}
    \caption{Participants’ responses: (a) mental demand, (b) physical demand, (c) engagement, and (d) overall user experience ranking. Solid lines indicate statistical significance with p $<$ 0.05, and dashed lines indicate p $<$ 0.1. The tables below display effect sizes for pairwise comparisons, with bold text highlighting the outperforming conditions.} 
    \label{fig:ratings}
\end{figure}

\textbf{\raisebox{-0.04\height}\unifiedB~enables faster transitions.}
We observed that the \unifiedT~consistently faster than the \separatedT~in both instructed and exploratory tasks (see~\autoref{fig:performance}-(a) and (b)). 
Performing within a unified environment enabled participants to complete tasks faster (five) and skip steps (four) by reducing the need to revisit target artifacts to recall information, particularly in navigational transitions.
As expected, transitions were facilitated by simply moving eyes or heads in the \unifiedT, whereas \separatedT~required additional steps (e.g., walking into portals) to switch workspaces~\cite{yang2020embodied}, increasing the total time.
Subjective ratings also confirmed that \unifiedT~provided better transition, with twelve in \unifiedT~preferring the transition experience. 
As one participant noted, ``I could skip steps in transitioning and save a lot of time (P12).'' 
Therefore, we observed that \unifiedT~was more efficient and faster, consistent with $H_{space-time}$.

\textbf{Low-effort transition within \raisebox{-0.045\height}\unifiedB~reduced the cognitive load.}
We observed that transitioning in \unifiedT~significantly lowered the mental load, as shown in~\autoref{fig:ratings}-(a). 
We believe the need to memorize and retrieve information was replaced by the ability to quickly transition and check information, resulting in more total transitions in the \unifiedT~than \separatedT~for both interactive and navigational transitions (see~\autoref{fig:performance}-(c), (d), and (e)).
Notably, we observed that the transition counts were distinct when assessing navigational transitions.
In our study, 6 participants in the \unifiedT~reported that \unifiedT~lowered mental demand.
One participant commented, ``I don't need to memorize much, as I don't need to spend more time on transitions (P4).''
Our findings partially align with previous studies~\cite{ahmad2023framework}, exploring how frequent information checking can impact mental load.
We believe these observations confirm both $H_{space-mental}$ and $H_{space-trans}$.

\textbf{\raisebox{-0.045\height}\unifiedB~potentially better-supported workflow tracking.}
We further observed that diminished capability in workflow tracking potentially introduces an additional burden to the mental load in the \separatedT~(see \autoref{fig:ratings}-(a)).
Tracking a workflow often involves identifying and recognizing previously encountered objects to reduce redundant processing~\cite{mancusi2023trackflow}. 
Within interactive transitions, we observed that almost all participants could directly navigate to the target cell in \unifiedT~to sync data artifacts.
We believe this is because participants could facilitate spatial information by displaying data artifacts near the target cell--- in front of the cell, as shown in \autoref{fig:vr_condition}-(b).
In contrast, we observed that facilitating spatial information in a spatially disconnected environment was impractical in \separatedT. 
This led to considerable time spent identifying target cells due to the need to adjust to a new workspace with sudden changes, making it harder to recall the cell's position and resulting in higher context-switching costs~\cite{yang2020embodied}.

\textbf{\raisebox{0.07\height}\vrB~potentially promotes provenance.}
In data science, provenance refers to documenting the origins, lineage, and processing history of data, which is essential for ensuring traceability and reliability~\cite{wang2015big}.
In our study, provenance played a key role in helping participants maintain their workflow by allowing them to track and revisit the history of their actions.
Interestingly, our results did not show a statistically significant difference between the \separatedT~and \unifiedT~in terms of the number of artifacts left or deleted (see~\autoref{fig:performance}-(f) and (g)).
While we did not confirm $H_{space-del\_left}$, 11 participants across both conditions expressed appreciation for the large display space.
This finding aligns with prior studies in embodied data transformation~\cite{in2023table}, where participants tended to keep more data tables in VR.
We believe this reflects the benefits of immersive environments, where larger display spaces make it easier for participants to view and interact with a greater number of artifacts.


\textbf{Limitations and Future Work.}
Although ICoN aims to facilitate seamless transitions, some improvements need to be discussed.
First, while the current low-code approach supports widely used packages like Pandas, Matplotlib, and Seaborn, expanding to additional libraries, such as SciPy~\cite{2020SciPy-NMeth}, and incorporating diverse plot types will enable a more comprehensive exploration of data. 
In addition, visual linking mechanisms, while effective in clarifying workflows~\cite{yu2016visflow}, sometimes obscure information due to excessive links, underscoring the need for optimization to highlight only relevant connections. 
Regarding scalability, the provided number of codes in the study aligned with typical lengths~\cite{in2024evaluating, ramasamy2023workflow}, but the exploration of larger computational notebooks in immersive spaces has not been extensively investigated.
Therefore, evaluating larger datasets is essential for accurately simulating real-world scenarios.

%% file: body/09-Conclusion.tex
\section{Conclusion}
\label{sec:conclusion}
In this study, we introduce ICoN, a system designed to facilitate a seamless transition between computational notebooks and embodied data science tasks within a unified environment.
We confirmed that ICoN enhances the transition experiences between computational notebooks and embodied data interaction workspaces.
Furthermore, unifying computational notebooks with embodied data science tasks helps reduce mental load, as users can quickly retrieve and reference information during their analysis.
However, we observed that participants may struggle with managing the workspace when dealing with a larger number of artifacts in complex data science tasks.
In summary, we believe ICoN illustrates an innovative immersive data science workspace by offering a novel approach to transition between code, data transformation, and visualization.